\documentclass[11pt]{article}
\usepackage{amsfonts,amsmath}
\usepackage[vcentermath]{youngtab}
\usepackage{young}
\def\bbt{\bibitem}
\def\be{\begin{equation}}
\def\en{\end{equation}}
\def\ber{\begin{eqnarray}}
\def\enr{\end{eqnarray}}
\def\nmb{ \nonumber\\}
\def\d{\partial}
\def\rbr{\rbrack}
\def\lbr{\lbrack}
\def\rbrc{\rbrace}
\def\lbrc{\lbrace}
\def\ov{\over }
\def\tld{\tilde}

\def\MTR{Manin triple }

\def\DLG{double Lie group }
\def\al{\alpha}
\def\bet{\beta}
\def\Tt{\Theta}
\def\tt{\theta}
\def\sgm{\sigma}

\def\gm{\gamma}
\def\Gm{\Gamma}
\def\im{\imath}

\def\lm{\lambda}

\def\dlt{\delta}
\def\om{\omega}
\def\Om{\Omega}

\def\trho{\tilde{\rho}}
\def\txi{\tilde{\xi}}

\begin{document}

\centerline{\bf Generalized K$\ddot{a}$hler Geometry and current algebras}
\centerline{\bf in classical N=2 superconformal WZW model.}
\vskip 1.5 true cm
\centerline{\bf S. E. Parkhomenko}
\centerline{Landau Institute for Theoretical Physics}
\centerline{142432 Chernogolovka,Russia}
\vskip 0.5 true cm
\centerline{spark@itp.ac.ru}
\vskip 0.5 true cm
\centerline{\bf Abstract}
\vskip 0.5 true cm
I examine the Generalized K$\ddot{a}$hler geometry of classical $N=(2,2)$ superconformal WZW model on a compact group and relate the right-moving and left-moving Kac-Moody superalgebra currents to the 
Generalized K$\ddot{a}$hler geometry data using biholomorphic gerbe formulation and Hamiltonian formalism. It is shown that canonical Poisson homogeneous space structure induced by the Generalized K$\ddot{a}$hler geometry of the group manifold is crucial to provide consistently $N=(2,2)$ superconformal $\sgm$-model with the Kac-Moody superalgebra symmetries. Then the biholomorphic gerbe geometry is used to prove that Kac-Moody superalgebra currents are globally defined. 
\vskip 10pt
 
\centerline{\bf Introduction.}

 It is well-known that conformal supersymmetric $\sgm$-models with extended supersymmetry play the role of building blocks in the construction
of realistic models of superstring compactification from 10 to 4 dimensions \cite{Gep}. The $\sgm$-model on the 6 -dimensional Calabi-Yau manifold is one of the important examples of the compactification. The Calabi-Yau manifold being a complex K$\ddot{a}$hler manifold is not accidental but caused by a close relation between the extended supersymmetry and K$\ddot{a}$hler geometry. In a more general case the background geometry may include apart from the metric an antisymmetric $B$-field. In that case the corresponding 2-dimensional supersymmetric $\sgm$-model have a second supersymmetry when the target-space has a bi-Hermitian geometry, known also as Gates-Hull-Ro$\check{c}$ek geometry \cite{GHR}. In this situation the target manifold contains two complex structures with a Hermitian metric with respect to each of the complex structures. Quite recently it has been shown in \cite{Gualt} that these set of geometric objects, metric, antisymmetric $B$-field and two complex structures antisymmetric with respect to the metric have a unified description in the context of Generalized K$\ddot{a}$hler (GK) geometry. It allowed to develop the GK geometry construction of $N=2$ Virasoro superalgbra for these models in the papers \cite{Z}, \cite{BLPZ}, \cite{Zlec}, \cite{HelZ}, \cite{HelZ1}.

 The $N=2$ supersymmetric WZW models on the compact groups \cite{SSTP}, \cite{P}, \cite{QFR} provide a large class of examples of exactly solvable quantum conformal $\sgm$-models whose targets supports simultaneously GK geometry causing the extended $N=(2,2)$-supersymmetry and affine Kac-Moody superalgebra structure ensuring the exact solution \cite{KacTod}, \cite{JF}. The GK geometry nature of the $N=(2,2)$-supersymmetry in these models has been studied in the series of works \cite{SevT}, \cite{RSSev}, \cite{HLRUZ}, 
\cite{SevSW}, \cite{SevST}. The relation between Generalized Complex geometry and Kac-Moody superalgebra symmetry has already been also discussed in papers \cite{HelZ}, \cite{HelZ1}, \cite{P1}. It has been noticed in particular in \cite{P1} that canonical Poisson homogeneous space structure induced by the GK geometry on $SU(2)\times U(1)$ is essential to provide consistently $N=(2,2)$ superconformal $\sgm$-model with Kac-Moody superalgebra symmetries. In this paper I consider this relation in more detailes and generalize the discussion from \cite{P1} for an arbitrary (even dimensional) compact group manifold.  

 In a more general context it would be important to see if there are GKG targets which allow the $W$-superalgabras conserved currents. Perhaps Kazama-Suzuki coset models \cite{KS1}-\cite{KS3} can be related to such targets according to the results presented in \cite{P3}, \cite{HS}. I hope the analysis developed in this paper appears to be helpfull for the GK geometry investigation in Kazama-Suzuki models. 


 The approach I follow is based on the papers \cite{Z}, \cite{BLPZ}, \cite{Zlec} where the Hamiltonian formalism of $N=1$ supersymmetric $\sgm$-model has been applied. It was shown there that the canonical formalism is very important to relate $N=(2,2)$ Virasoro superalgebra symmetries of the $\sgm$-model with the GK geometry of the target space.  In case of $N=(2,2)$ WZW model the canonical variables play yet another role. They can be naturally fixed to have Poisson-Lie geometrical sense.

 In section 1 I apply biholomorphic gerbe formulation of GK geometry (bi-Hermitian geometry) discovered in \cite{HLRUZ}   to the $N=(2,2)$ WZW model on a compact even dimensional group and introduce certain complex coordinate covering of the group manifold. The coordinates of the covering are given by the action of certain complex group which is Drinfeld's dual to the WZW model compact group. For each complex coordinate chart the canonical variables of the model are determined from the WZW action. Then it is shown that the canonical variables corresponding to different coordinate charts are consistent on the intersections and generate a sheaf of Poisson Vertex superalgebras twisted by WZW 3-form. Thus biholomorphic gerbe structure is naturaly appears and essential at this point. In section 2 I find for some fixed coordinate chart, the left-moving and right-moving conserved currents in terms of the canonical variables and certain GK geometry Poisson byvector which is know in the theory of Integrable Systems as Semenov-Tian-Shansky Poisson byvector. Then the geometry behind the expressions for the conserved currents is explained in terms of the Poisson-Lie geometry, canonical variables and covering introduced in section 1. It allows to calculate the Poisson brackets between the currents and prove the Kac-Moody superalgebra relations they satisfy. The Poisson homogeneous space geometry on the group manifold determined by Semenov-Tian-Shansky Poisson byvector is crucial for the calculation. Using the twisted Poisson Vertex algebra and biholomorphic gerbe structure it is proved in section 3 that the Kac-Moody superalgebra currents found in section 2 are globally defined. In section 4 we discuss the $N=(2,2)$ supersymmetric Sugawara construction and explain the expressions for the Virasoro superalgebra currents as well as their Poisson brackets in terms of Poisson-Lie structures induced by the GK geometry. Concluding remarks are collected in the section 5.


\vskip 10pt

\centerline{\bf 1. Canonical variables and biholomorphic gerbe geometry}
\centerline{\bf in $N=(2,2)$ supersymmetric WZW model.}

{\bf 1.1.} In the papers \cite{P},\cite{QFR} a correspondence between extended
supersymmetric WZW models and finite-dimensional Manin triples was established. By the definition ~\cite{Drinf}, a \MTR $({\bf p},{\bf g_{+}},{\bf g_{-}})$
consists of a Lie algebra ${\bf p}$, with nondegenerate invariant inner product
$<,>$ and isotropic Lie subalgebras ${\bf g_{\pm}}$ such that
${\bf p}={\bf g_{+}}\oplus {\bf g_{-}}$ as a vector space. 
There is a one to one correspondence \cite{P} between a complex Manin triple endowed with antilinear involution which conjugates isotropic subalgebras $\tau: {\bf g_{\pm}}\rightarrow {\bf g_{\mp}}$ and
a complex structure on a real Lie algebra of the compact group that ensures connection between Manin triple construction of  $N=2$   Virasoro superalgebra currents of \cite{P},\cite{QFR} and approach \cite{SSTP} based on the complex structure on the Lie algebra. Namely, for an arbitrary real Lie algebra ${\bf g}$ with the complex structure $J$
the complexification ${\bf g^{\mathbb{C}}}$ has the Manin triple
structure $({\bf g}^{\mathbb{C}},{\bf g_{+}},{\bf g_{-}})$ where the isotropic subalgebras are the $\pm\im$-eigenspaces
of $J$ while ${\bf g}$ is the fixed point set of the antilinear involution $\tau$. For some fixed
orthonormal basis $\{e^{a}, e_{a}$, $a= 1,...,d\}$ in algebra
${\bf g}^{\mathbb{C}}$ so that $\{e^{a}\}$ is a basis in ${\bf g_{-}}$, $\{e_{a}\}$
is a basis in ${\bf g_{+}}$ the commutation relations of ${\bf g^{\mathbb{C}}}$
have the form
\ber  
\lbr e^{a},e^{b}\rbr=\phi^{ab}_{c}e^{c},     
\
\lbr e_{a},e_{b}\rbr=\phi_{ab}^{c}e_{c},     
\nmb
\lbr e^{a},e_{b}\rbr=\phi_{bc}^{a}e^{c}-\phi^{ac}_{b}e_{c}. 
\label{1.commutator}
\enr
In what follows I will often use common notation $f^{ij}_{k}$ for the structure constants of the Lie algebra 
${\bf g}^{\mathbb{C}}$. There is another Lie algebra structure on the vector space ${\bf g}^{\mathbb{C}}$ which is given by the direct sum ${\bf g}_{+}\oplus {\bf g}_{-}$ of the isotropic subalgabras. For any vectors 
$a,b\in {\bf g}^{\mathbb{C}}$ the new Lie algebra brackets $\lbr a, b\rbr_{J}$ are given by
\ber
\lbr a, b\rbr_{J}={1\ov 2}(\lbr Ja,b\rbr+\lbr a, Jb\rbr)
\label{1.Jcommut}
\enr
This Lie algebra structure which we denote by ${\bf g}^{\mathbb{C}}_{J}$ is very important in the theory of integrable systems \cite{SemTian}.

 The Lie group version
of the triple $({\bf g}^{\mathbb{C}},{\bf g_{+}},{\bf g_{-}})$ is the \DLG $({\bf G}^{\mathbb{C}},{\bf G}_{+},{\bf G}_{-})$
~\cite{SemTian}, \cite{LuW}. 
The real compact Lie group ${\bf G}$ is extracted
from its complexification by the involution (Hermitian conjugation) $\tau$:
\ber
{\bf G}= \lbrc g\in {\bf G}^{\mathbb{C}}|\tau (g)=g^{-1}\rbrc       
\label{1.realG}
\enr
In this case there are two complex structures $J_{R}$ and $J_{L}$ on the compact group, where $J_{R}$ is generated by the right translations on the group (left-invariant complex structure) and $J_{L}$ is generated by the left translations (right-invariant complex structure). Bi-invariant metric on the group is Hermitian w.r.t. both of the complex structures so that we have bi-Hermitian geometry \cite{GHR} on the group manifold ${\bf {G}}$.
 
 Exponentiating the commutation relations (\ref{1.commutator}) one can find that ${\bf G}^{\mathbb{C}}={\bf G}_{+}{\bf G}_{-}={\bf G}_{-}{\bf G}_{+}$ so that each element $g\in {\bf G}^{\mathbb{C}}$
admits two decompositions
\ber
g= g_{+}g^{-1}_{-}= {\tld g}_{-}{\tld g}^{-1}_{+},  
\label{1.decomp}
\enr
where ${\tld g}_{\pm}$ are dressing transformed
elements of $g_{\pm}$ ~\cite{LuW}.
In view of (\ref{1.realG}) and (\ref{1.decomp}) the
element $g$ from ${\bf G}^{\mathbb{C}}$ belongs to the compact group ${\bf G}$ iff
\ber
\tau (g_{\pm})= {\tld g}^{-1}_{\mp}      
\label{1.unitar}
\enr
These equations mean that there is a left and right actions of the complex groups ${\bf G}_{\pm}$ on ${\bf G}$ \cite{SemTian} so the elements of ${\bf G}$ can be parametrized
by the elements from the complex group ${\bf G}_{+}$ (or ${\bf G}_{-}$).

 More generaly one has to consider the set $W$ of classes ~\cite{AlMal}
\ber
{\bf {G}}^{\mathbb{C}}= \bigcup_{[w]\in {\bf{W}}} {\bf{G}}_{+}w{\bf{G}}_{-}
\label{1.Bruhat}
\enr
It will be more convenient to rewrite this decomposition in
another form
\ber
{\bf {G}}^{\mathbb{C}}= \bigcup_{[w]\in {\bf{W}}}w{\bf G}^{w}_{+}{\bf G}_{-}:
\
h=wh^{w}_{+}h_{-}^{-1}
\label{1.Bruhat1}
\enr
where
\ber
h^{w}_{+}\in{\bf G}^{w}_{+}= {\bf G}_{+}\cap w^{-1}{\bf G}_{+}w, \ h_{-}\in {\bf G}_{-}
\label{1.Bruhat2}
\enr
The decomposition (\ref{1.Bruhat1}) can be converted into the covering of the group by the open subsets if we multiply the Bruhat classes from the left by the special elements from ${\bf G}_{-}$ generating transverse directions:
\ber
{\bf {G}}^{\mathbb{C}}=\bigcup_{[w]\in {\bf{W}}}{\bf G}^{-}_{w}w{\bf G}^{w}_{+}{\bf G}_{-}
\label{1.Bruhat3}
\enr
where
\ber
{\bf G}^{-}_{w}=\lbrc u_{-}\in {\bf G}_{-}|w^{-1}u_{-}w\in {\bf G}_{+}\rbrc
\label{1.Bruhat4}
\enr
If we assume that the representatives $w$ can be choosen to belong to the compact group
\ber
\tau(w)=w^{-1}
\label{1.wunitary}
\enr
the covering (\ref{1.Bruhat3}) will induce the covering of the compact group by the charts which are isomorphic to the complex group ${\bf {G}}_{+}$:
\ber
{\bf{G}}=\bigcup_{[w]\in {\bf{W}}}{\bf{G}}_{w},
\nmb
{\bf{G}}_{w}=\lbrc wg_{+}g_{-}^{-1}|g_{\pm}\in {\bf{G}}_{\pm}, \ \tau (g_{\pm})= {\tld g}^{-1}_{\mp} \rbrc
\label{1.LBrchart}
\enr
where ${\tld g}_{\pm}$ are determined from the equation (\ref{1.decomp}).

 It is not difficult to see that the covering defines $J_{R}$-holomorphic diffeomorphisms
\ber
\varphi_{w}:{\bf{G}}_{w}\rightarrow{\bf{G}}_{+}
\label{1.wmap}
\enr
so that 1-forms $dg_{+}g_{+}^{-1}$ considered as the forms on ${\bf {G}}_{w}$ are the holomorphic w.r.t. the complex structure $J_{R}$ \cite{P2}.

 On the equal footing another covering can be used
\ber
{\bf{G}}=\bigcup_{[w]\in {\bf{W}}}\tld{{\bf{G}}}_{w},
\nmb
\tld{{\bf{G}}}_{w}=\lbrc \tld{g}_{-}\tld{g}_{+}^{-1}w|\tld{g}_{\pm}\in {\bf{G}}_{\pm}, \ 
\tau (\tld{g}_{\pm})=g^{-1}_{\mp} \rbrc
\label{1.RBrchart}
\enr
where $g_{\pm}$, $\tld{g}_{\pm}$ satisfy (\ref{1.decomp}).
This covering defines $J_{L}$-holomorphic diffeomorphisms
\ber
\tld{\varphi}_{w}:\tld{{\bf{G}}}_{w}\rightarrow{\bf{G}}_{+}
\label{1.mapw}
\enr
so that 1-forms $d\tld{g}_{+}\tld{g}_{+}^{-1}$ considered as the forms on $\tld{{\bf {G}}}_{w}$ are the holomorphic w.r.t. the complex structure $J_{L}$ \cite{P2}.


{\bf 1.2.}  
 The super world-sheet of the $WZW$ model is parametrized by the light-cone
even coordinates
$\sgm_{\pm}$, and odd coordinates $\tt_{\pm}$. The super-derivatives are given by
\ber
D_{\pm}={\d\ov\d\tt^{\pm}}+\tt^{\pm}\d_{\pm}, \ \d_{\pm}={\d\ov \d \sgm_{1}}\pm{\d\ov \d\sgm_{0}}
\label{1.lconeD}
\enr
The superfield $g(\sgm_{\pm},\tt^{\pm})$ takes values in the compact group ${\bf G}$. The action of the model \cite{swzw} is given by
\ber
S= -k(\int d^{2}\sgm d^{2} \tt(<g^{-1}D_{+}g,g^{-1}D_{-}g>   
\nmb
         -\int d^{2}\sgm d^{2}\tt dt
          <g^{-1}\frac{\d g}{\d t},\lbrc g^{-1}D_{-}g,g^{-1}D_{+}g\rbrc>)
\label{1.action}
\enr 
The classical equations of motion are nothing else but the conservation low equations:
\ber
 D_{-}(g^{-1}D_{+}g)=D_{+}(D_{-}gg^{-1})= 0.
\label{1.meq}
\enr

 For any coordinate chart  from (\ref{1.LBrchart}) or (\ref{1.RBrchart}) the action (\ref{1.action}) can be rewritten in the form of the supersymmetric $\sgm$-model action by
the Polyakov-Wiegman formula ~\cite{PW}. For the the chart ${\bf{G}}_{w}$ where $J_{R}$-complex coordinates 
$x^{\mu}=(x^{\al},\bar{x}^{\al})$ are introduced the action is:
\ber
S={1\ov 2}\int d^{2}\sgm d^{2} \tt E^{w}_{ij}\rho^{i}_{+}\rho^{j}_{-}
\label{1.action1}
\enr
Here, the matrix $E^{w}_{ij}$ depends on the superfields $X^{\mu}$ corresponding to the coordinates 
$x^{\mu}$. 
The superfields $\rho^{i}_{\pm}=(\rho^{a}_{\pm},\rho^{\bar{a}}_{\pm})=(\rho)^{i}_{\mu}D_{\pm}X^{\mu}$ are the world-sheet restrictions of the holomorphic
$\rho^{a}=Tr(e^{a}dg_{+}g_{+}^{-1})$ and anti-holomorphic $\rho^{\bar{a}}=\bar{\rho}^{a}$ 1-forms w.r.t $J_{R}$. 
The target space metric $g^{w}_{ij}$ and skew-symmetric $B$-field $B^{w}_{ij}$ can be read off from $E^{w}_{ij}$:
\ber
g^{w}_{ij}={1\ov 2}(E^{w}_{ij}+E^{w}_{ji}), \ B^{w}_{ij}=-{1\ov 2}(E^{w}_{ij}-E^{w}_{ji})
\label{1.gB}
\enr
The metric and WZW 3-form do not depend on the coordinate chart
\ber
g^{w}_{ij}=g_{ij}, \ H={1\ov 2}dB^{w}
\label{1.metricHWZW}
\enr
  
 The $\sgm$-model Lagrangian can be rewritten in the following form \cite{P2}:
\ber
E^{w}_{ij}=\im\Om^{w}_{ik}(J_{R})^{k}_{j}
\label{1.wSTSform}
\enr
where $\Om^{w}$ is the image of the Semenov-Tian-Shansky symplectic form \cite{AlMal} 
\ber
\Om=
{k\ov 4}(<dg_{+}g_{+}^{-1}\hat{,} dg_{-}g_{-}^{-1}>+<d\tld{g}_{+}\tld{g}_{+}^{-1}\hat{,} d\tld{g}_{-}\tld{g}_{-}^{-1}>)
\label{1.STSform}
\enr
under the map $\varphi_{w}$.
Due to (\ref{1.LBrchart}), (\ref{1.RBrchart}) Semenov-Tian-Shansky symplectic form is real. 

 On the equal footing the charts from (\ref{1.RBrchart}) can be used to write the action in
$J_{L}$-complex coordinates $\tld{x}^{\mu}=(\tld{x}^{\al},\bar{\tld{x}}^{\al})$, associated to 
the chart $\tld{{\bf{G}}}_{w}$:
\ber
S={1\ov 2}\int d^{2}\sgm d^{2} \tt \tld{E}^{w}_{ij}\trho^{i}_{+}\trho^{j}_{-}
\label{1.tildaction1}
\enr
The superfields $\trho^{i}_{\pm}=(\trho^{a}_{\pm},\trho^{\bar{a}}_{\pm})$ are the world-sheet restrictions of the holomorphic
$\trho^{a}=Tr(e^{a}d\tld{g}_{+}\tld{g}_{+}^{-1})$ and anti-holomorphic $\trho^{\bar{a}}=\bar{\trho}^{a}$ 
1-forms w.r.t $J_{L}$. In this case we have
\ber
\tld{E}^{w}_{ij}=\im\tld{\Om}^{w}_{ik}(J_{L})^{k}_{j}
\label{1.STSformw}
\enr
where $\tld{\Om}^{w}$ is the image of the Semenov-Tian-Shansky symplectic form (\ref{1.STSform}) under the map $\tld{\varphi}_{w}$.

{\bf 1.3.}  Now we find the canonically conjugated momentum and Hamiltonian following the analisis of \cite{BLPZ}.
Let us define new set of odd world-sheet coordinates and derivatives
\ber
\tt^{+}={1\ov\sqrt{2}}(\tt^{0}+\tt^{1}), \ \tt^{-}={1\ov\sqrt{2}}(\tt^{0}-\tt^{1}),
\nmb
D_{0}={\d\ov\d\tt^{0}}+\tt^{0}\d_{1}+\tt^{1}\d_{0},
\
D_{1}={\d\ov\d\tt^{1}}+\tt^{1}\d_{1}+\tt^{0}\d_{0}
\label{1.newsuperderiv}
\enr
so that $D_{0}^{2}=D_{1}^{2}=\d_{1}$, $D_{0}D_{1}+D_{1}D_{0}=2\d_{0}$ and $\d_{1}$ ($\d_{0}$) is the derivative along the space coordinate $\sgm^{1}$ (time coordinate $\sgm^{0}$) on the world-sheet.

 Let us consider first the chart ${\bf {G}}_{1}$ with the coordinates $x^{\mu}$. The canonically conjugated momentum to the field $X^{\mu}$ as well as the Hamiltonian density ${\cal H}$ are found by integrating out $\tt^{0}$-variable in the action (\ref{1.action1}) \cite{P1}:
\ber
X^{*}_{\mu}=g_{\mu\nu}D_{0}X^{\nu}+B_{\mu\nu}D_{1}X^{\nu}
\label{1.momentumB}
\enr
\ber
{\cal H}=g_{\mu\nu}\d_{1} X^{\mu}D_{1}X^{\nu}+g^{\mu\nu}I_{\mu}D_{1}I_{\nu}-
\Gm^{\lm}_{\mu\nu}g^{\nu\sgm}I_{\lm}I_{\sgm}D_{1}X^{\mu}-
\nmb
H_{\mu\nu\lm}g^{\lm\sgm}D_{1}X^{\mu}D_{1}X^{\nu}I_{\sgm}+{1\ov 3}H^{\mu\nu\lm}I_{\mu}I_{\nu}I_{\lm}
\label{1.Hamilton}
\enr
where $I_{\mu}=X^{*}_{\mu}-B_{\mu\nu}DX^{\nu}$ and
\ber
\Gm^{\eta}_{\nu\lm}={1\ov 2}g^{\eta\mu}(g_{\mu\nu,\lm}+g_{\lm\mu,\nu}-g_{\nu\lm,\mu})
\label{1.Christof}
\enr
are the Christofell symbols of the Levi-Civita connection.

 The canonical Poisson super-brackets for the superfields $X^{\mu}(Z)$, $X^{*}_{\nu}(Z)$ living on the super-circle with the super-space coordinate $Z=(\sgm^{1}, \tt^{1})$ (in what follows the superscript index 1 will be ommitted) are given by
\ber
\lbrc X^{*}_{\mu}(Z_{1}), X^{\nu}(Z_{2})\rbrc=-\lbrc X^{\nu}(Z_{2}), X^{*}_{\mu}(Z_{1})\rbrc=
\dlt^{\nu}_{\mu}\dlt(Z_{1}-Z_{2})=
\nmb
\dlt^{\nu}_{\mu}\dlt(\sgm_{1}-\sgm_{2})(\tt_{1}-\tt_{2})
\label{1.PBcanon}
\enr

 One can find similarly the canonically conjugated momentum $Y^{*}_{\mu}$ to the field $Y^{\mu}$ using the chart ${\bf {G}}_{w}$ with coordinates $y^{\mu}$ and action (\ref{1.action1}). Then on the intersection 
${\bf{G}_{1w}}={\bf{G}}_{1}\bigcap {\bf{G}}_{w}$ the following gluing rule takes place
\ber
Y^{\nu}(Z)=y^{\nu}(X^{\mu}(Z)),
\nmb
Y^{*}_{\nu}={\d x^{\mu}\ov \d y^{\nu}}(X^{*}_{\mu}+(B^{w}-B^{1})_{\mu\lm}D_{1}X^{\lm})
\label{1.momentrans}
\enr
It is clear that similar relations are true for an arbitrary intersection ${\bf{G}_{w'w}}$. 

 In view of (\ref{1.metricHWZW}) the set of 2-forms $B^{w}$ defines the set of 1-forms $A_{w'w}$ determined on the intersections ${\bf G}_{w'w}$ such that
\ber
B^{w}-B^{w'}=dA^{w'w},
\label{1.gerbe1}
\enr
When $k\in \mathbb{Z}$ this data define gerbe with connection \cite{Bril}, \cite{HLRUZ}. Taking into account 
(\ref{1.wmap}), (\ref{1.mapw}), (\ref{1.wSTSform}), (\ref{1.STSformw}) one can see that we are in the sitution of theorem 
from paper \cite{HLRUZ} where very nice description of GK geometry in terms of locally defined symplectic forms determining the biholomorphic gerbe was found. In our case this set of forms is given by the Semenov-Tian-Shansky symplectic forms $\Om^{w}$, $\tld{\Om}^{w}$.

 The canonical Poisson brackets are related closely to the Courant brackets (as well as the Dorfman brackets) \cite{Hel}. Let us consider the currents $V(Z)=v^{i}X^{*}_{i}(Z)+\om_{i}D_{1}X^{i}(Z)$, $Q(Z)=q^{j}X^{*}_{j}+\lm_{j}D_{1}X^{j})(Z)$,
where $(v(X),\om(X))$, $(q(X),\lm(X))$ are the superfields considering as the sections of direct sum $TG\oplus T^{*}G$ of tangent and cotangent bundles over the group manifold. Then
\ber
\lbrc V(Z_{1}),Q(Z_{2})\rbrc=D_{Z_{1}}\dlt(Z_{1}-Z_{2})2<V,Q>(Z_{2})+
\nmb
\dlt(Z_{1}-Z_{2})(\lbr V,Q\rbr_{c}+D_{Z_{2}}<V,Q>)(Z_{2})
\label{1.Courant}
\enr
where the bracket $\lbr V,Q\rbr_{c}=\lbr(v,\om),(q,\lm)\rbr_{c}=\lbr v,q\rbr+(Lie(v)\lm-Lie(q)\om)-{1\ov 2}d(\lm(v)-\om(q))$ is nothing else but
Courant brackets while the expression $\lbr V,Q\rbr_{c}+d<V,Q>$ is Dorfman brackets and $<V,Q>={1\ov 2}(\lm(v)+\om(q))$ is the natural pairing. Then the gluing rules (\ref{1.momentrans}) comes from the twisted Courant algebroid structure on $TG\oplus T^{*}G$. In fact we have also a sheaf of twisted Poisson Vertex algebras \cite{Hel}, \cite{Bril}, \cite{EHKZ} whose sections over ${\bf G}_{w}$ are generated by the canonical variables $X^{\mu}(Z)$, $X^{*}_{\mu}(Z)$ and their super-derivatives along the super-circle variable $Z$.

\vskip 10pt
\centerline{\bf 2. GK and Poisson-Lie geometry of Kac-Moody superalgebra currents.}

{\bf 2.1.} Two copies of Kac-Moody superalgebra symmeties generated by the left-moving $L=D_{+}gg^{-1}$ and right-moving $R=g^{-1}D_{-}g$ currents is the main feature of the superconformal WZW model.  These two algebras determine the dynamics and $N=2$ supersymmetry of the model completely due to the generalized Sugawara construction 
\cite{SSTP}, \cite{P}, \cite{QFR} (see also \cite{Getz}) and make the model completely solvable. 

 The expressions for the left-moving and right-moving currents written in terms of GK geometry data of the group manifold has been found in \cite{P1} (in a more general context similar formulas have also been
obtained in \cite{AlStr}, \cite{Fedya}). Namely, the classical conserved left-moving currents are given by
\ber
L^{i}=-{k\ov 2}(P(\rho^{i})+\im J_{R}\rho^{i})=-{k\ov 2}(P^{ij}\rho^{*}_{j}+\im (J_{R})^{i}_{k}\rho^{k})
\label{2.LPoisson}
\enr
where $\rho^{*}_{j}=(\rho^{-1})^{\mu}_{j}X^{*}_{\mu}$ is the conjugated to $\rho^{j}$ momentum field.
The right-moving currents is convenient to write in the coordinates $\tld{{\bf{G}}}_{1}$ 
(notice that $\tld{{\bf{G}}}_{1}={\bf{G}}_{1}$ because of (\ref{1.decomp})) 
\ber
R^{i}=-{k\ov 2}(P(\tld{\rho}^{i})-\im J_{L}\tld{\rho}^{i})=-{k\ov 2}(P^{ij}\trho^{*}_{j}-\im (J_{L})^{i}_{k}\trho^{k})
\label{2.RPoisson}
\enr
where $\tld{\rho}^{i}=(Tr(e^{a}D_{1}\tld{g}_{+}\tld{g}_{+}),Tr(e_{a}D_{1}\tld{g}_{+}\tld{g}_{+}))$
are the super-circle restrictions of holomorphic and anti-holomorphic 1-forms w.r.t. the complex structutre $J_{L}$.
It is implied here that the coordinates $\tld{x}^{i}$ have been introduced on $\tld{{\bf{G}}}_{1}$ so that
$\tld{X}^{i}$, $\tld{X}^{*}_{i}$ are canonically conjugated superfields and $\tld{\rho}^{*}_{i}$ is conjugated to
$\tld{\rho}^{i}$.
The Semenov-Tian-Shansky Poisson byvector $P=\Om^{-1}$ in the expressions above is given by the complex structures and metric \cite{SemTian} (see also \cite{HelZ1}):
\ber
P=-\im(J_{L}+J_{R})g^{-1}
\label{2.STSPoisson}
\enr

 It worth to explain at this point the Poisson-Lie geometry behind the expressions (\ref{2.LPoisson}), (\ref{2.RPoisson}) which appears to be crucial to provide the Kac-Moody superalgebra Poisson brackets for the currents.
Recall that by definition the right dressing (left deressing) vector field \cite{LuW} on the compact group ${\bf G}$ considered as a Poisson manifold with Poisson byvector $P=\Om^{-1}$ is given by substitution of the right invariant (left invariant) 1-form on ${\bf G}$ into the Poisson byvector $P$. Then the first term of the formula (\ref{2.LPoisson}), ((\ref{2.RPoisson})) comes because of the $\rho^{*}_{i}$, ($\trho^{*}_{i}$) is the loop group version of the right dressing (left deressing) vector field on ${\bf G}_{1}$. In other words, the first term of 
(\ref{2.LPoisson}), ((\ref{2.RPoisson})) is nothing else but the loop group version of the relation between the left translation (right translation) vector field and right dressing (left dressing) vector field on the Poisson manifold 
${\bf G}_{1}$. The forms 
$\rho^{i}$ ($\trho^{i}$) coming from the group ${\bf G}_{+}$ play the role of differentials of local Hamiltonians for the left translations (right translations) so that $\varphi_{1}$ ($\tld{\varphi}_{1}$) is a momentum map. 
Thus, the Drinfeld's dual group \cite{Drinf} ${\bf G}_{+}$ determines the canonical momentum variables by the dressing action while the canonical coordinates associated to the coordinates on ${\bf G}_{+}$ play the role of Hamiltonians for the Kac-Moody superalgebra symmetries. 

 The second terms from (\ref{2.LPoisson}) and (\ref{2.RPoisson}) follow unambigously from the definitions of conjugated momentum fields $X^{*}_{\mu}$ and $\tld{X}^{*}_{\mu}$. Being proportional to $k$ these terms define the levels of Kac-Moody superalgebras as we will see calculating the Poisson brackets of the currents in the next subsection. So the critical level $k=0$ the Kac-Moody superalgebra currents are given entirely by the Semenov-Tian-Shansky Poisson byvector.

\leftline {\bf 2.2. Proposition.} The currents (\ref{2.LPoisson}), (\ref{2.RPoisson}) generate two copies of mutually commuting Kac-Moody superalgebras at level $k$.

\leftline{\it Proof.} It is proving by the Poisson super-brackets calculation. To this end, notice that the currents 
(\ref{2.LPoisson}) are the fuctions of even $X^{\mu}(Z)$ and odd $X^{*}_{\mu}(Z)$ variables whose Poisson brackets are given by (\ref{1.PBcanon}). The currents (\ref{2.RPoisson}) are the fuctions of even $\tld{X}^{\mu}(Z)$and odd $\tld{X}^{*}_{\mu}(Z)$ variables whose Poisson brackets are given obvoiusly also by (\ref{1.PBcanon}) (where the untilded symbols replaced by tilded ones). During the proof we will use the notation $D$ for the super-derivative along the super-circle while the symbols $D_{1}$ or $D_{2}$ will be used to denote the $\dlt$-function super-derivatives w.r.t. the $\dlt$-function arguments $Z_{1}$ or $Z_{2}$.

 Let us calculate first the Poisson super-brackets for the left-moving currents:
\ber
{4\ov k^{2}}\lbrc L^{i}(Z_{1}),L^{j}(Z_{2})\rbrc=
\nmb
\lbrc P^{\mu\nu}\rho^{i}_{\mu}X^{*}_{\nu}(Z_{1}), P^{\lm\rho}\rho^{j}_{\lm}X^{*}_{\rho}(Z_{2})\rbrc+
\im(J_{R})^{i}_{k}\lbrc\rho^{k}_{\nu}DX^{\nu}(Z_{1}),P^{\lm\rho}\rho^{j}_{\lm}X^{*}_{\rho}(Z_{2})\rbrc+
\nmb
\im(J_{R})^{j}_{n}\lbrc P^{\mu\nu}\rho^{i}_{\mu}X^{*}_{\nu}(Z_{1}),\rho^{n}_{\lm}DX^{\lm}(Z_{2})\rbrc
\label{2.LLPB1}
\enr
The first brackets calculation gives
\ber
\lbrc P^{\mu\nu}\rho^{i}_{\mu}X^{*}_{\nu}(Z_{1}), P^{\lm\rho}\rho^{j}_{\lm}X^{*}_{\rho}(Z_{2})\rbrc=
-\dlt (Z_{1}-Z_{2})P^{\sgm\nu}(\lbrc\rho^{i},\rho^{j}\rbrc_{P})_{\nu}X^{*}_{\sgm}(Z_{2})
\label{2.LLPB2}
\enr
where $\lbrc\rho^{i},\rho^{j}\rbrc_{P}$ is determined by 
\ber
(\lbrc\rho^{i},\rho^{j}\rbrc_{P})_{\nu}=P^{\mu\lm}_{,\nu}\rho^{i}_{\mu}\rho^{j}_{\lm}+
P^{\mu\lm}\rho^{i}_{\nu,\mu}\rho^{j}_{\lm}+P^{\mu\lm}\rho^{i}_{\mu}\rho^{j}_{\nu,\lm}
\label{2.formsPbrack}
\enr
But
\ber
\lbrc\rho^{i},\rho^{j}\rbrc_{P}=-{2\ov k}f^{ij}_{k}\rho^{k}
\label{2.Liebracket1}
\enr
as it follows from the discussion after the formula (\ref{2.STSPoisson}). Hence
\ber
\lbrc P^{\mu\nu}\rho^{i}_{\mu}X^{*}_{\nu}(Z_{1}), P^{\lm\rho}\rho^{j}_{\lm}X^{*}_{\rho}(Z_{2})\rbrc=
\dlt (Z_{1}-Z_{2}){2\ov k}f^{ij}_{k}P^{\sgm\nu}\rho^{k}_{\nu}X^{*}_{\sgm}(Z_{2})
\label{2.LLPB3}
\enr

 Calculating the second brackets from (\ref{2.LLPB1}) we obtain
\ber
\im(J_{R})^{i}_{k}\lbrc\rho^{k}_{\nu}DX^{\nu}(Z_{1}),P^{\lm\rho}\rho^{j}_{\lm}X^{*}_{\rho}(Z_{2})\rbrc=
\nmb
\im D_{1}\dlt(Z_{1}-Z_{2})(J_{R})^{i}_{k}P^{\lm\rho}\rho^{j}_{\lm}\rho^{k}_{\nu}(Z_{2})
\nmb
-{\im\ov 2}\dlt(Z_{1}-Z_{2})(J_{R})^{i}_{k}P^{\lm\nu}\phi^{k}_{nm}\rho^{j}_{\lm}
(\rho^{n}_{\nu}\rho^{m}_{\sgm}-\rho^{n}_{\sgm}\rho^{m}_{\nu})DX^{\sgm}(Z_{2})
\label{2.LLPB4}
\enr
where $\phi^{k}_{nm}=(\phi_{ab}^{c}, \bar{\phi}_{ab}^{c}=\phi^{ab}_{c})$ are the structure constants of the 
Lie algebra ${\bf g}^{\mathbb{C}}_{J}$ (see (\ref{1.Jcommut})). They come from the Maurer-Cartan equations for 1-forms $dg_{+}g_{+}^{-1}$,
$d\bar{g}_{+}\bar{g}_{+}^{-1}$ on the group manifold ${\bf G}_{+}$.

The third brackets from (\ref{2.LLPB1}) are
\ber
\im(J_{R})^{j}_{n}\lbrc P^{\mu\nu}\rho^{i}_{\mu}X^{*}_{\nu}(Z_{1}),\rho^{n}_{\lm}DX^{\lm}(Z_{2})\rbrc=
\nmb
-\im D_{2}\dlt(Z_{1}-Z_{2})(J_{R})^{j}_{n}P^{\mu\nu}\rho^{i}_{\mu}\rho^{n}_{\nu}+
\nmb
\im\dlt(Z_{1}-Z_{2})(J_{R})^{j}_{n}(\lbrc\rho^{i},\rho^{n}\rbrc_{\lm}-
{1\ov 2}\phi^{i}_{km}P^{\mu\nu}\rho^{n}_{\nu}(\rho^{k}_{\mu}\rho^{m}_{\lm}-\rho^{k}_{\lm}\rho^{m}_{\mu}))DX^{\lm}(Z_{2})
\label{2.LLPB5}
\enr
Collecting all these results we obtain
\ber
{4\ov k^{2}}\lbrc L^{i}(Z_{1}),L^{j}(Z_{2})\rbrc=
\nmb
\im D_{1}\dlt(Z_{1}-Z_{2})(P(\rho^{j},J_{R}\rho^{i})+P(\rho^{i},J_{R}\rho^{j}))+
\nmb
\dlt(Z_{1}-Z_{2})(-{2\ov k}f^{ij}_{k}P(\rho^{k})-\im{2\ov k}(J_{R})^{j}_{n}f^{in}_{k}\rho^{k}-
\nmb
{\im\ov 2}\phi^{k}_{nm}(J_{R})^{i}_{k}P(\rho^{j},\rho^{n})\rho^{m}+
{\im\ov 2}\phi^{k}_{nm}(J_{R})^{i}_{k}P(\rho^{j},\rho^{m})\rho^{n}+
\nmb
{\im\ov 2}\phi^{i}_{nm}(J_{R})^{j}_{k}P(\rho^{k},\rho^{n})\rho^{m}-
{\im\ov 2}\phi^{i}_{nm}(J_{R})^{j}_{k}P(\rho^{k},\rho^{m})\rho^{n})
\label{2.LLPB6}
\enr
To simplify the expression the following relations have to be used
\ber
P(J_{R}\rho^{j},\rho^{n})-P(\rho^{j},J_{R}\rho^{n})=-2\im g^{jn}
\label{2.Pgrelation}
\enr
\ber
\phi^{k}_{nm}(J_{R})^{i}_{k}=(J_{R})^{k}_{n}\phi^{i}_{km}
\label{2.Jfirelation}
\enr
The first one follows from (\ref{2.STSPoisson}), (\ref{1.gB}) and (\ref{1.wSTSform}).
The second relation follows because of $J_{R}$ acts on the Lie algebra ${\bf g}^{\mathbb{C}}_{J}$ by $\pm\im$ multiplication.
Hence we obtain
\ber
{4\ov k^{2}}\lbrc L^{i}(Z_{1}),L^{j}(Z_{2})\rbrc=
\nmb
2g^{ij}D_{1}\dlt(Z_{1}-Z_{2})+
\nmb
\dlt(Z_{1}-Z_{2})({4\ov k^{2}}f^{ij}_{k}L^{k}+\im{2\ov k}f^{ij}_{k}(J_{R})^{k}_{n}\rho^{n}-
\im{2\ov k}f^{ik}_{n}(J_{R})^{j}_{k}\rho^{n}+{4\ov k}\phi^{i}_{nm}g^{nj}\rho^{m})(Z_{2})
\label{E.LLPB7}
\enr

 Now we use (\ref{1.Jcommut})
\ber
\phi^{i}_{nm}g^{nj}=-{\im\ov 2}(f^{ij}_{k}(J_{R})^{k}_{m}-f^{ik}_{m}(J_{R})^{j}_{k})
\label{2.Jcommutator}
\enr
which finally gives the expected result 
\ber
\lbrc L^{i}(Z_{1}),L^{j}(Z_{2})\rbrc=
k\hat{g}^{ij}D_{1}\dlt(Z_{1}-Z_{2})+\dlt(Z_{1}-Z_{2})f^{ij}_{k}L^{k}(Z_{2})
\label{2.LLPB8}
\enr
The metric tensor $\hat{g}_{ij}$ here is a rescaled version of the metric tensor $g_{ij}$
$\hat{g}_{ij}={2\ov k}g_{ij}$ so it does not depend on $k$.

 To calculate the Poisson brackets between the right-moving currents one has to repeat all steps above chaging $J_{R}\rightarrow -J_{L}$,
$\rho\rightarrow \trho$ and taking into account that
\ber
\lbrc\trho^{i},\trho^{j}\rbrc_{P}={2\ov k}f^{ij}_{k}\trho^{k}
\label{2.RRPB1}
\enr
since the Hamiltonians $\trho^{i}$ determine the right translations on ${\bf G}$.
The result is
\ber
\lbrc R^{i}(Z_{1}),R^{j}(Z_{2})\rbrc=
-k\hat{g}^{ij}D_{1}\dlt(Z_{1}-Z_{2})-\dlt(Z_{1}-Z_{2})f^{ij}_{k}R^{k}(Z_{2})
\label{2.RRPB2}
\enr

 To calculate the Poisson brackets between left-moving and right-moving currents notice first that
\ber
\lbrc \rho^{i},\trho^{j}\rbrc_{P}=0
\label{2.LRPB1}
\enr
Then
\ber
{4\ov k^{2}}\lbrc L^{i}(Z_{1}),R^{j}(Z_{2})\rbrc=
-\im D_{1}\dlt(Z_{1}-Z_{2})(P(\rho^{i},J_{L}\trho^{j})+P(J_{R}\rho^{i},\trho^{j})(Z_{2})+
\nmb
-\im\dlt(Z_{1}-Z_{2})\phi_{km}^{i}(P(\rho^{m},J_{L}\trho^{j})+P(J_{R}\rho^{m},\trho^{j}))\rho^{k}(Z_{2})
\label{2.LRPB2}
\enr
But
\ber
P(\rho^{i},J_{L}\trho^{j})+P(J_{R}\rho^{i},\trho^{j})=0
\label{2.LRPB3}
\enr
due to (\ref{2.STSPoisson}). Hence
\ber
\lbrc L^{i}(Z_{1}),R^{j}(Z_{2})\rbrc=0
\label{2.LRPB4}
\enr

 It finishes the proof.

\vskip 10pt
\centerline{\bf 3. The Kac-Moody superalgebra currents are globally defined.}

 The currents (\ref{2.LPoisson}), (\ref{2.RPoisson}) and their Poisson algebra brackets have been determined 
in the chart ${\bf G}_{1}$. They can also be determined for any other chart ${\bf G}_{w}$ as well as for $\tld{{\bf G}}_{w}$. As it follows from (\ref{1.action1}), (\ref{1.tildaction1}),
they are given by the expressions similar to (\ref{2.LPoisson}), (\ref{2.RPoisson}) but written in terms of local canonical fields $Y^{\mu}(Z)$, $Y^{*}_{\mu}(Z)$ ($\tld{Y}^{\mu}(Z)$, $\tld{Y}^{*}_{\mu}(Z)$) introduced in the chart 
${\bf G}_{w}$ ($\tld{{\bf G}}_{w}$). 
So the question is how they glue to each other on the intersections ${\bf G}_{w'w}$. We are going to show that the currents considered as the sections of sheaf of twisted Poisson Vertex algebras are defined globally.

 Let us pick up an element $w\in {\bf{W}}$ and consider the chart ${\bf{G}}_{w}$ with the the complex coordinates 
$y^{\mu}$. The WZW action functional written for this chart takes the form 
\ber
S_{wzw}[wh_{+}h_{-}^{-1}]=
{1\ov 2}\int d^{2}\sgm d^{2}\Tt E^{w}_{ij}\xi^{i}_{+}\xi^{j}_{-},
\
E^{w}_{ij}=\im \Om^{w}_{ik}(J_{R})^{k}_{j},
\label{3.waction}
\enr
where $\xi_{\pm}=D_{\pm}h_{+}h_{+}^{-1}$, $h_{+}\in {\bf G}_{+}$. Let $(L_{w})^{i}(Z)$ be the left-moving currents determined on ${\bf G}_{w}$. They are given by
\ber
(L_{w})^{i}=-{k\ov 2}(w)^{i}_{j}((P_{w})^{\mu\nu}\xi^{j}_{\mu}Y^{*}_{\nu}+\im(J_{R})^{j}_{k}\xi^{k}_{\nu}D_{1}Y^{\nu})
\label{3.wLcurrent}
\enr
where $P_{w}=(\Om^{w})^{-1}$ is Semenov-Tian-Shansky Poisson byvector defined in this chart, and the matrix $w^{i}_{j}$ is determined 
from $w^{-1}e_{j}w=(w^{-1})^{i}_{j}e_{i}$.

 Let us consider the intersection ${\bf G}_{1w}={\bf{G}}_{1}\bigcap {\bf{G}}_{w}$. 
There are two systems of complex coordinates. The coordinates $y^{\mu}$ are related to the chart ${\bf{G}}_{w}$ and the coordinates $x^{\mu}$ are related to the
chart ${\bf{G}}_{1}$. It is easy to see that the map 
\ber
\dot{w}=\varphi_{w}\circ \varphi_{1}^{-1}: g_{+}\rightarrow h_{+}, \ y^{\mu}=y^{\mu}(x)
\label{3.Rholw}
\enr
is $J_{R}$-holomorphic diffeomorphism because it is given by the left shift on ${\bf G}$ by the element $w$.

\leftline {\bf {Proposition L.}}

 On the intersection ${\bf{G}_{1w}}={\bf{G}}_{1}\bigcap {\bf{G}}_{w}$:
$(L_{w})^{i}=L^{i}$.

\leftline{\it Proof.} (During the proof we will ommit the subscript $1$ in the notation of super-derivative along the coordinate $\tt^{1}$ and write $D$ instead of $D_{1}$).
Due to (\ref{1.momentrans}) we have
\ber
(w^{-1})^{i}_{j}(L_{w})^{j}=
-{k\ov 2}((P_{w})^{\mu\nu}\xi^{i}_{\mu}Y^{*}_{\nu}+\im(J_{R})^{i}_{k}\xi^{k}_{\nu}DY^{\nu})=
\nmb
-{k\ov 2}\xi^{i}_{\mu}(y){\d y^{\mu}\ov\d x^{\gm}}{\d x^{\gm}\ov\d y^{\sgm}}(P_{w})^{\sgm\nu}(y(x))
{\d x^{\al}\ov\d y^{\nu}}(X^{*}_{\al}+(w^{*}B^{w}-B^{1})_{\al\bet}DX^{\bet})-
\nmb
\im{k\ov 2}\xi^{i}_{\mu}(y)(J_{R})^{\mu}_{\lm}{\d y^{\lm}\ov\d x^{\bet}}DX^{\bet}
\label{3.Ltrans1}
\enr
The complex structure $J_{R}$ is invariant under the left shifts on the group:
\ber
{\d x^{\gm}\ov\d y^{\sgm}}{\d y^{\lm}\ov\d x^{\bet}}(J_{R}(y(x)))^{\sgm}_{\lm}=(J_{R}(x))^{\gm}_{\bet}
\label{2.trans2}
\enr
Hence
\ber
-{2\ov k}(w^{-1})^{i}_{j}(L_{w})^{j}=\xi^{i}_{\mu}{\d y^{\mu}\ov\d x^{\gm}}(\dot{w}_{*}^{-1}P_{w})^{\gm\al}(x)X^{*}_{\al}+
\im\xi^{i}_{\mu}{\d y^{\mu}\ov\d x^{\gm}}(J_{R}(x))^{\gm}_{\bet}DX^{\bet}+
\nmb
\xi^{i}_{\mu}{\d y^{\mu}\ov\d x^{\gm}}(w_{*}^{-1}P_{w})^{\gm\al}(x)(w^{*}B^{w}-B^{1})_{\al\bet}DX^{\bet}
\label{3.trans3}
\enr
where
\ber
(\dot{w}_{*}^{-1}P_{w})^{\gm\al}(x)={\d x^{\gm}\ov\d y^{\sgm}}{\d x^{\al}\ov\d y^{\mu}}(P_{w}(y(x)))^{\sgm\mu}
\nmb
(\dot{w}^{*}B^{w})_{\gm\al}(x)={\d y^{\mu}\ov\d x^{\gm}}{\d y^{\nu}\ov\d x^{\al}}(B^{w}(y(x)))_{\mu\nu}
\label{3.trans4}
\enr
Written by the basic forms $\rho^{i}$ the expression (\ref{3.trans3}) takes the form
\ber
-{2\ov k}(w^{-1})^{i}_{j}(L_{w})^{j}=\xi^{i}_{n}(\dot{w}_{*}^{-1}P_{w})^{nm}(x)\rho^{*}_{m}+
\nmb
\im\xi^{i}_{n}(J_{R})^{n}_{m}\rho^{m}+
\xi^{i}_{n}(\dot{w}_{*}^{-1}P_{w})^{nm}(\dot{w}^{*}B^{w}-B^{1})_{ml}\rho^{l}
\label{3.trans5}
\enr

But
\ber
\xi^{i}_{n}(\dot{w}_{*}^{-1}P_{w})^{nm}(x)\rho^{*}_{m}=(w^{-1})^{i}_{j}(P_{1}(x))^{jm}\rho^{*}_{m}\Leftrightarrow
\nmb
(\dot{w}^{-1})_{*}(P_{w}(\xi^{i}))=P_{1}((w^{-1})^{i}_{j}\rho^{j})
\label{3.equivar}
\enr
because of the vector fields $P_{1}(\rho^{i})$, $P_{w}(\xi^{i})$ are the left translations on ${\bf G}_{1w}$ and the map $\dot{w}$ acts on them as a differential of the left shift by $w$.


 Therefore we obtain
\ber
-{2\ov k}(w^{-1})^{i}_{j}(L_{w})^{j}=(w^{-1})^{i}_{j}(P_{1}^{jm}\rho^{*}_{m}+\im(J_{R})^{j}_{m}\rho^{m})+
\nmb
\im(\xi^{i}_{n}-(w^{-1})^{i}_{n})(J_{R})^{n}_{m}\rho^{m}+
(w^{-1})^{i}_{n}P_{1}^{nl}(\dot{w}^{*}B_{w}-B_{1})_{lm}\rho^{m}
\label{3.trans7}
\enr
Now one uses the relations
\ber
(\dot{w}^{*}B^{w}-B^{1})_{lm}=-\im (\dot{w}^{*}\Om^{w}-\Om^{1})_{lk}(J_{R})^{k}_{m}
\label{3.trans8}
\enr
which easelly follow from (\ref{2.STSPoisson}), (\ref{1.gB}), (\ref{1.wSTSform}). It gives
\ber
-{2\ov k}(w^{-1})^{i}_{j}(L_{w})^{j}=-{2\ov k}(w^{-1})^{i}_{j}L^{j}+
\nmb
\im(\xi^{i}_{n}-(w^{-1})^{i}_{n})(J_{R})^{n}_{m}\rho^{m}-
\im(w^{-1})^{i}_{n}(P_{1}^{nl}(\dot{w}^{*}\Om^{w})_{lk}-\dlt^{n}_{k})(J_{R})^{k}_{m}\rho^{m}
\label{3.trans9}
\enr
Therefore the left currents are globally defined iff
\ber
\xi^{i}_{n}((\dot{w}^{-1})_{*}P_{w})^{nl}=(w^{-1})^{i}_{n}P_{1}^{nl}
\label{3.trans10}
\enr
The last relation is nothing else but (\ref{3.equivar}), so the proposition follows.

 To prove the corresponding statement for the right-moving currents one needs to consider the second covering and take the chart $\tld{\bf{G}}_{w}$ with the coordinates $\tld{y}^{\mu}$. The WZW action functional written for this chart takes the form
\ber
S_{wzw}[\tld{h}_{-}\tld{h}_{+}^{-1}w]=
{1\ov 2}\int d^{2}\sgm d^{2}\Tt \tld{E}^{w}_{ij}\txi^{i}_{+}\txi^{j}_{-},
\
\tld{E}^{w}_{ij}=\im \tld{\Om}^{w}_{ik}(J_{L})^{k}_{j}, 
\nmb 
\tld{g}_{ij}={1\ov 2}(\tld{E}^{w}_{ij}+\tld{E}^{w}_{ji}), 
\ \tld{B}^{w}_{ij}={1\ov 2}(\tld{E}^{w}_{ij}-\tld{E}^{w}_{ji})
\label{E.actionw}
\enr
where $\txi=D_{1}\tld{h}_{+}\tld{h}_{+}^{-1}$.

 We determine the canonical coordinates $\tld{Y}^{\mu}(\sgm,\tt)$ and canonical momenta $\tld{Y}^{*}_{\mu}(\sgm,\tt)$ from this action as well as the currents 
\ber
(R_{w})^{i}=
-{k\ov 2}(w^{-1})^{i}_{j}((\tld{P}_{w})^{\mu\nu}\txi^{j}_{\mu}\tld{Y}^{*}_{\nu}-\im(J_{L})^{j}_{k}\txi^{k}_{\nu}D\tld{Y}^{\nu})
\label{E.Rcurrentw}
\enr

 Let us consider the intersection $\tld{{\bf G}}_{1w}=\tld{{\bf{G}}}_{1}\bigcap \tld{{\bf{G}}}_{w}$. 
There are two systems of complex coordinates $\tld{y}^{\mu}$ and $\tld{x}^{\mu}$ there. The map 
\ber
\dot{\tld{w}}=\tld{\varphi}_{w}\circ \tld{\varphi}_{1}^{-1}: \tld{g}_{+}\rightarrow \tld{h}_{+}, \ \tld{y}^{\mu}=\tld{y}^{\mu}(\tld{x})
\label{3.Rholw}
\enr
is $J_{L}$-holomorphic diffeomorphism because it is given by the right shift on ${\bf G}$.

Therefore one can analogously prove 

\leftline {\bf{Proposition R.}}

 On the intersection $\tld{{\bf{G}}}_{1w}=\tld{{\bf{G}}}_{1}\bigcap \tld{{\bf{G}}}_{w}$:
$(R_{w})^{i}=R^{i}$.
 
 The Propositions L and R are generalized straightforwardly for the intersections ${\bf{G}}_{w'w}$ and $\tld{{\bf{G}}}_{w'w}$. Thus we conclude that the currents $L^{i}$ and $R^{i}$ are defined globally.

\vskip 10pt
\centerline{\bf 4. $N=(2,2)$ supersymmetric Sugawara construction and Poisson-Lie group structure.}

  In this section we discuss Poisson-Lie structure underlying $N=(2,2)$ supersymmetric Sugawara construction.
	
 For the case of $\sgm$-model on a general GK geometry target space, the $N=(2,2)$ superalgebra Virasoro currents construction has been developed in the papers \cite{Z}, \cite{BLPZ}, \cite{Zlec}, \cite{HelZ}, \cite{HelZ1}. The generalized mutually commuting complex structures ${\bf J}_{1,2}$ on the target manifold play the central role in the construction. They are defined in the direct sum of tangent and cotangent bundles of the manifold and can be represented as
the following matrices 
\ber
{\bf J}_{1}={1\ov 2}\left(\begin{array}{cc}
1  & 0 \\
B & 1 \\
\end{array}\right)\left(\begin{array}{cc}
J_{L}+J_{R}     & -\om^{-1}_{L}+\om^{-1}_{R} \\
\om_{L}-\om_{R} & -J^{T}_{L}-J^{T}_{R}       \\
\end{array}\right)
\left(\begin{array}{cc}
1  & 0 \\
-B  & 1 \\
\end{array}\right)
\nmb
{\bf J}_{2}={1\ov 2}\left(\begin{array}{cc}
1  & 0 \\
B & 1 \\
\end{array}\right)\left(\begin{array}{cc}
J_{L}-J_{R}     & -\om^{-1}_{L}-\om^{-1}_{R} \\
\om_{L}+\om_{R} & -J^{T}_{L}+J^{T}_{R}       \\
\end{array}\right)
\left(\begin{array}{cc}
1  & 0 \\
-B  & 1 \\
\end{array}\right)
\label{4.GCstruct}
\enr
where 
\ber
\om_{L,R}=gJ_{L,R}, \ \om^{-1}_{L,R}=-J_{L,R}g^{-1}
\label{4.symplforms}
\enr

 The left-moving $K_{L}$ and right-moving $K_{R}$ $U(1)$-supercurrents of the $N=(2,2)$ Virasoro superalgebra are given by \cite{BLPZ}, \cite{SevST}
\ber
K_{L}=\im<(DX,X^{*}),({\bf J}_{1}+{\bf J}_{2})(DX,X^{*})>=
\nmb
\im(\om_{L}^{-1})^{\mu\nu}(X^{*}_{\mu}+(g-B)_{\mu\lm}DX^{\lm})
(X^{*}_{\nu}+(g-B)_{\nu\sgm}DX^{\sgm})
\nmb
K_{R}=\im<(DX,X^{*}),({\bf J}_{1}-{\bf J}_{2})(DX,X^{*})>=
\nmb
-\im(\om_{R}^{-1})^{\mu\nu}(X^{*}_{\mu}-(g+B)_{\mu\lm}DX^{\lm})(X^{*}_{\nu}-(g+B)_{\nu\sgm}DX^{\sgm})
\label{4.U1currents}
\enr

 For the case of group manifold target space they can be expressed on the chart ${\bf G}_{1}$ in terms of the left-moving and right-moving currents (\ref{2.LPoisson}), (\ref{2.RPoisson}) \cite{HelZ} (see also \cite{P1})
\ber
K_{L}={\im\ov 2k}(\hat{\om}_{L})_{ij}L^{i}L^{j},
\
K_{R}=-{\im\ov 2k}(\hat{\om}_{R})_{ij}R^{i}R^{j}
\label{4.classicKLR}
\enr
where $\hat{\om}_{L,R}=\hat{g}J_{L,R}$. They are defined globally and coincide with the classical limit of the currents found in \cite{SSTP}, \cite{P} (see also \cite {Getz}, \cite{QFR}).

 The symplectic forms $\om_{L,R}$ entering into (\ref{4.classicKLR}) are important in the construction of Poisson-Lie group structures on the manifold ${\bf G}$. Indeed, the Poisson byvector (\ref{2.STSPoisson}) provides the group manifold ${\bf G}$ with the structure of Poisson homogeneous space where the Poisson-Lie group ${\bf G}$ endowed with Poisson byvector
\ber
P_{DS}=-\im(J_{L}-J_{R})g^{-1}
\label{4.PDS}
\enr
acts on the Poisson homogeneous space ${\bf G}$ by the Poisson maps \cite{SemTian}. This can be traced by the calculation
of Poisson brackets:
\ber
\lbrc L^{i}(Z_{1}),K_{L}(Z_{2})\rbrc=
\nmb
\im D_{1}\dlt(Z_{1}-Z_{2})(J_{L})^{i}_{n}L^{n}(Z_{2})+\dlt(Z_{1}-Z_{2}){1\ov k}\varphi_{nk}^{i}L^{n}L^{k}(Z_{2})
\label{4.LKPB}
\enr
\ber
\lbrc R^{i}(Z_{1}),K_{R}(Z_{2})\rbrc=
\nmb
+\im D_{1}\dlt(Z_{1}-Z_{2})(J_{R})^{i}_{n}R^{n}(Z_{2})+\dlt(Z_{1}-Z_{2}){1\ov k}\varphi_{nk}^{i}R^{n}R^{k}(Z_{2})
\label{4.RKPB}
\enr
The $\dlt$-function contributions in (\ref{4.LKPB}), (\ref{4.RKPB}) come from the Lie derivatives of 
$\om_{L,R}$ under the left and right translations on ${\bf G}$ correspondingly, but 
$\varphi_{nk}^{i}$ are the structure constants of the Drinfeld's dual Lie algebra 
${\bf g}^{\mathbb{C}}_{J}$. Therefore the $\dlt$-function contributions are nothing else but the infinitesimal version of the Poisson action condition of the Poisson-Lie group ${\bf G}$ on the Poisson homogeneous space ${\bf G}$ \cite{LuW}.
 
 Due to (\ref{4.U1currents}) the $\dlt$-function derivative terms in (\ref{4.LKPB}), (\ref{4.RKPB}) mean that Kac-Moody superalgebra currents are eigenvectors w.r.t. each of the generalized complex structures (\ref{4.GCstruct}). 
Since the generalized complex structures commute to each other the direct sum of the left-moving and right-moving Kac-Moody superalgebras is decomposed into the direct sum of four $\pm\im$-eigensubalgebras \cite{HelZ1}.

 The Poisson brackets of $U(1)$-supercurrents give the components $\Gm_{L}$, $\Gm_{R}$ of stress-energy supercurrent
\ber
\lbrc K_{L}(Z_{1}),K_{L}(Z_{2})\rbrc=
\dlt(Z_{1}-Z_{2})\Gm_{L}(Z_{2}),
\nmb
\Gm_{L}=({1\ov k}\hat{g}_{jn}DL^{j}L^{n}+{1\ov 3k^{2}}\hat{g}_{ij}\hat{g}_{nm}f^{jm}_{k}L^{i}L^{n}L^{k})(Z_{2})
\label{4.LKKPB}
\enr
\ber
\lbrc K_{R}(Z_{1}),K_{R}(Z_{2})\rbrc=
\dlt(Z_{1}-Z_{2})\Gm_{R}(Z_{2}),
\nmb
\Gm_{R}=-({1\ov k}\hat{g}_{jn}DR^{j}R^{n}+{1\ov 3k^{2}}\hat{g}_{ij}\hat{g}_{nm}f^{jm}_{k}R^{i}R^{n}R^{k})
\label{4.RKKPB}
\enr
where the integrability of the complex structures $J_{L,R}$ has been used to get the 3-currents terms.
These terms are the Schouten brackets of byvectors $\om_{L,R}^{-1}$ with themselves. Thus they are the anomalies in the corresponding Jacoby identities and only these contributions left at the critical level $k=0$.

 By the direct calculations one can show that the Hamiltonian of the $\sgm$-model (\ref{1.Hamilton}) obtained in the canonical formalism is expressed in terms of the stress-energy supertensor $\Gm_{L,R}$:
\ber
H={1\ov 2}\int d\sgm^{1} d\tt^{1} (\Gm_{L}(Z)-\Gm_{R}(Z))
\label{4.Hamilt}
\enr
This expression follows from the more precise relations
\ber
\Gm_{L}={1\ov 2}({\cal H}-{\cal P}), \ \Gm_{R}=-{1\ov 2}({\cal H}+{\cal P})
\label{4.PH}
\enr
for the Hamiltonian density (\ref{1.Hamilton}) and momentum desity
\ber
{\cal P}=DX^{\mu}DX^{*}_{\mu}+\d X^{\mu}X^{*}_{\mu}
\label{4.P}
\enr
written by the canonical variables from the chart ${\bf G}_{1}$. 

 To obtain (\ref{4.PH}) one needs to use (\ref{4.LKKPB}), (\ref{4.RKKPB}), (\ref{2.LPoisson}), (\ref{2.RPoisson}) as well as the fact that left translation vector fields and right translation vector fields on group manifold ${\bf G}$ are covariantly constant w.r.t. the corresponding connections $\nabla_{L,R}$ with torsion. It gives the relations between the connection coeficients of $\nabla_{L,R}$ and the derivatives of the Poisson byvector components. For example 
\ber
\Om_{\nu\sgm}P^{\sgm\mu}_{,\lm}+\Gm_{\nu\lm}^{\mu}+g^{\mu\sgm}H_{\sgm\nu\lm}=0
\label{4.Lcovconst}
\enr
takes place for the connection $\nabla_{L}$ in the coordinate chart ${\bf G}_{1}$. While 
\ber
\tld{\Om}_{\nu\sgm}\tld{P}^{\sgm\mu}_{,\lm}+\tld{\Gm}_{\nu\lm}^{\mu}-\tld{g}^{\mu\sgm}\tld{H}_{\sgm\nu\lm}=0
\label{4.Rcovconst}
\enr
takes place for the connection coefficients of $\nabla_{R}$ written in the coordinate chart $\tld{{\bf G}}_{1}$.

 It easy to see from (\ref{4.Hamilt}) that the currents $L^{i}$, $R^{i}$ defined on the super-circle are the left-moving and right-moving indeed. At the same time it follows from the formulas (\ref{4.LKPB}), 
(\ref{4.RKPB}) and the comment below them that the left-moving and right-moving Kac-Moody superalgebras
are also $\pm$-eigensubalgebras w.r.t. the product structure operator ${\bf J}_{1}{\bf J}_{2}$. Hence, the product structure is nothing else but the holomorphic-anti-holomorphic factorization of the $N=(2,2)$ WZW model and the Hamiltonian conserves this structure.

\vskip 10pt
\centerline{\bf 5. Conclusion.} 

 In this paper I discussed the relation between GK geometry causing $N=(2,2)$ Virasoro superalgebra symmetries and
the Kac-Moody superalgebra conserved currents ensuring the exact solution of supesymmetric WZW model on a compact even dimensional group. 
It has been shown that Poisson homogeneous space structure which is given by Semenov-Tian-Shansky Poisson byvector is crucial to privide the Kac-Moody superalgebra symmetries of the model.
The canonical formalism we used in the discussion is natural to relate $N=(2,2)$ Virasoro superalgebra symmetries with the GK geometry of the target space. But in case of $N=(2,2)$ WZW model the canonical variables play yet another role.
They can be determined locally by the biholomorphic gerbe geometry and the moment maps from the target space into the Drinfeld's dual complex group ${\bf G}_{+}$ in such a way that the canonical coordinates become the Hamiltonians for the Kac-Moody superalgebra action at the critical level while the canonical momenta are given by the dressing vector fields on the target space. These data define a sheaf of twisted Poisson Vertex algebras. This structure was used to prove that Kac-Moody superalgebra currents are defined globaly. 
It is shown that Sugawara construction of $N=(2,2)$ Virasoro superalgebra is based essentialy on the underlying Poisson homogeneous space geometry of the model also. 
 
 There are some questions which left beyond the scope of the paper but would be interesting to consider.

1) What is Poisson-Lie geometric sense of generalized K$\ddot{a}$hler potential? How to describe the pure spinors and chiral ring states from the point of view of Poisson-Lie geometry of the model?
The papers \cite{HLRUZ}, \cite{Z1}, \cite{ABM} certainly be helpful in this concern.

2) It would be interesting to see if the Kazama-Suzuki models \cite{KS1}-\cite{KS3} can be considered as a GK geometry targets with $W$-algebra of consereved currents. Due to the paper \cite{P3} it is natural to expect that the Poisson homogeneous space structure will be crucial also to provide $W$-algebra symmetries in these models.
 
3) Quantization of the geometric structures considered in the paper is probably the most intresting problem. It has already been discussed in a general context of $N=(2,2)$ $\sgm$-models in paper \cite{EHKZ}. The authors of the paper proposed in particular to consider the chiral de Rham complex \cite{MSV} as a formal canonical quantization of non-linear $\sgm$-model so that the sheaf of Poisson Vertex algebras is its classical limit. In case of quantum $N=(2,2)$ WZW model it would be interesting to see whether the quantization of sheaf of twisted Poisson Vertex algebras could reproduce the results known from the bootstrap approach. It is well known \cite{Drinf} that Poisson-Lie groups and Poisson homogeneous spaces are the classical limits of the quantum groups so that the Poisson-Lie structures discussed in this article may appear to be helpful in this concern.  




\vfill
\end{document}